\documentclass[conference,10pt]{IEEEtran}
\usepackage{cite}      
\usepackage{graphicx}  
\usepackage{url}
\usepackage{times}
\usepackage{amsmath}   
\usepackage{pdflscape}
\usepackage{balance}
\usepackage{amsfonts,amsthm,amssymb}
\usepackage{mathtools}
\usepackage{nicefrac}
\usepackage[none]{hyphenat}
\usepackage{array}
\usepackage{fontenc}
\usepackage{color}
\usepackage{etoolbox}
\usepackage{bm}
\usepackage{comment}
\usepackage{amsmath,amsthm,amssymb, amsfonts}
\usepackage{amssymb,mathtools}
\usepackage[normalem]{ulem}
\usepackage{nomencl}
\usepackage{accents}

\newlength{\dhatheight}

\usepackage{algorithm,algpseudocode}
\usepackage{xparse}
\makeatletter
\algnewcommand{\LineComment}[1]{\State  \(\triangleright\) #1 \hfill~}
\algrenewcommand{\LineComment}[1]{$\triangleright$ #1}

\makeatother
\begin{document}

\setlength{\textfloatsep}{0pt}
\setlength{\intextsep}{10pt plus 2pt minus 2pt}

\title{ Radar Operating Metrics and Network Throughput for Integrated Sensing and Communications in Millimeter-wave Urban Environments}

 \author{
	\IEEEauthorblockN{Akanksha~Sneh and   Shobha~Sundar~Ram\\
 Indraprastha Institute of Information Technology Delhi, New Delhi 110020 India\\
 \{akankshas, shobha\}@iiitd.ac.in}\

 }

\maketitle

\begin{abstract}
Millimeter wave integrated sensing and communication (ISAC) systems are being researched for next-generation intelligent transportation systems. Here, radar and communication functionalities share a common spectrum and hardware resources in a time-multiplexed manner. 
The objective of the radar is to first scan the angular search space and detect and localize mobile users/targets in the presence of discrete clutter scatterers. Subsequently, this information is used to direct highly directional beams toward these mobile users for communication service. The choice of radar parameters such as the radar duty cycle and the corresponding beamwidth are critical for realizing high communication throughput. In this work, we use the stochastic geometry-based mathematical framework to analyze the radar operating metrics as a function of diverse radar, target, and clutter parameters and subsequently use these results to study the network throughput of the ISAC system. The results are validated through Monte Carlo simulations.
\end{abstract}

\begin{IEEEkeywords}
integrated sensing and communication, stochastic geometry, radar operating metrics, Poisson point process
\end{IEEEkeywords}

\section{Introduction}
Over the last few years, there has been tremendous progress in the research and development of assisted and autonomous driving. However, detecting and recognizing objects remains an ongoing challenge for avoiding accidents and improving safety. Traditional object detection relies on sensors like lidar, cameras, radars, and infrared sensors. But radar is the only sensor that works well in poor visibility and bad weather. Currently, millimeter-wave automotive (mmW) radars are the preferred technology for detecting objects since they have wide bandwidths (around 2-4 GHz) and high-range resolution \cite{peng2022modern}. Concurrently, next-generation intelligent transportation systems are being developed to support various vehicle-to-everything (V2X) communication frameworks, including vehicle-to-vehicle (V2V), vehicle-to-infrastructure (V2I), and vehicle-to-pedestrian (V2P) communications. The main goal of these frameworks is to promote the sharing of road and vehicle data for environmental monitoring, collision avoidance, and pedestrian detection. 

Recently, active research has been directed towards integrating radar sensing and communication functionalities on the same platform for reducing spectral congestion and hardware costs \cite{deng2013interference,li2017joint,martone2017spectrum,hassanien2016signaling,hessar2016spectrum}.  Conventional standalone frequency modulated continuous wave (FMCW) waveforms currently used in automotive radars \cite{alaee2019discrete} are not optimized for communications. Similarly, existing communication protocols such as the dedicated short-range communications \cite{kenney2011dedicated}, cellular V2X \cite{wang2018cellular} and device-to-device V2X communications \cite{asadi2014survey} operate below 6 GHz and have very low bandwidths for supporting radar remote sensing. Instead, the IEEE 802.11ad protocol at the unlicensed 60 GHz mmW band has been identified as a viable solution for integrated sensing and communications (ISAC) because of the wide bandwidth and suitable signal structure \cite{choi2016millimeter}. As a result, there has been substantial research into the research and development of 802.11ad-based ISAC radar for both radar sensing and communications in recent works for automotive systems \cite{kumari_ieee_2018,duggal2020doppler,muns_beam_2019,10319100,sneh2023radar}. 

In all of these works, the radar and communication functionalities share the spectrum and hardware resources in a time-multiplexed manner where the radar within the access point/base station first detects and localizes mobile users/targets in the field of view. Subsequently, the communications commence between the multifunctional access point and the mobile user. The choice of the duty cycle between the duration of radar operation and subsequent communications is critical. For example, a longer radar exploration time implies that a narrow radar beam is used to scan the search space. This results in higher gain and, thereby detection of weaker targets. However, a longer radar exploration time limits communication service times to shorter durations. On the other hand, a shorter radar exploration time necessitates a broader radar beam for scanning the same search space. This results in lower overall gain in the radar link metric resulting in the missed detections of weak targets. However, as a result, the remaining detected targets can access longer communication service times. The overall throughput of the ISAC system is thus a function of the radar duty cycle. In prior work, the duty cycle between the radar and communication functionalities was optimized for maximum throughput for specific signal-to-clutter and noise ratios (SCNR) using stochastic geometry (SG) under line-of-sight conditions \cite{ram2022optimization}. In this work, we extend the work by examining the radar operating metrics - probabilities of false alarm and detection - and network throughput under noise and clutter-limited conditions in more complex propagation environments.

SG provides a mathematical framework for evaluating performance metrics of spatial stochastic processes that mimic Poisson point process distributions, eliminating the need for computationally intensive system simulations or time-consuming field measurements.
In prior art, \cite{al2017stochastic} used SG tools to characterize the statistics of radar interference. 
Here, the distribution of interfering automotive radars in an urban road scenario was modeled as a homogeneous Poisson point process. The authors in \cite{ram2020estimating,ram2022estimation,singhal2023leo} modeled the discrete clutter scatterers encountered in monostatic and bistatic radar scenarios as a Poisson point process and 
estimated the radar detection metrics. Then, they utilized this framework to optimize pulse radar parameters for maximizing the radar detection metrics \cite{ram2021optimization}.

In this work, we estimate the radar operating metrics for ISAC using SG tools for an mmW urban environment. Further, we study the effect of radar, target, and clutter parameters on the probability of false alarms and network throughput. Our paper is organized as follows. We discuss the radar signal model in Section~\ref{sec:Signal Model} followed by the estimation of radar operating metrics in Section~\ref{sec:ROM}. Further, we discuss our ISAC model in Section~\ref{sec:System Model} followed by the estimation of network throughput. Lastly, we present the numerical results in Section~\ref{sec: Numerical Results} and the conclusion in Section~\ref{sec:Conclusion}.

\emph{Notation:} In this paper, bold font denotes the random variables, while regular font denotes constants and realizations of a random variable.
\section{Radar Signal Model}
\label{sec:Signal Model}
\begin{figure}[htbp]
    \centering   
    \includegraphics[scale=0.5]{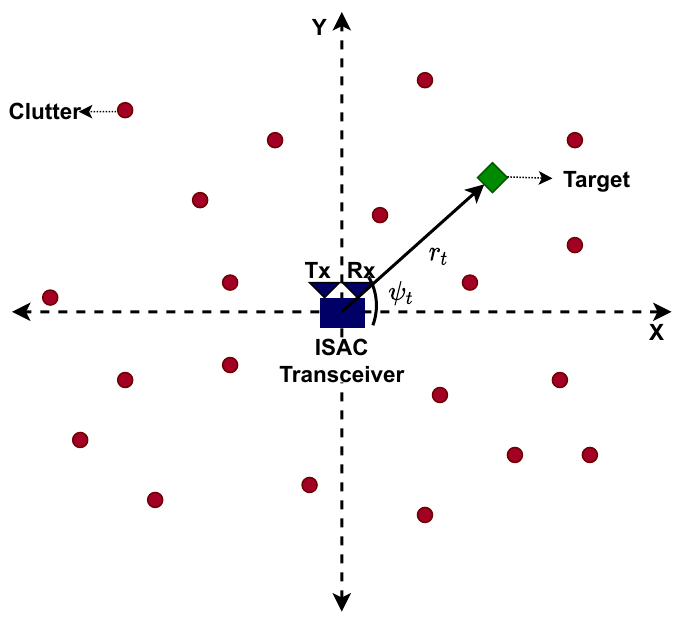}
    \caption{System model of integrated sensing and communication framework. Here the dual-functional monostatic radar/communication unit is assumed to be located at the origin. The target indicated by the green diamond is located within the main lobe. There are discrete clutter scatterers represented by the red dots distributed over the entire angular search space $\Omega$.}
    \label{fig:sys_model}
\end{figure}

We consider a dual radar-communication system located at the origin. When the system operates as a monostatic radar, it transmits signals of power $P_{tx}$ and receives the scattered signals from both a target represented by a green diamond in Fig.~\ref{fig:sys_model} and discrete clutter scatterers in the channel represented as red dots in Fig.~\ref{fig:sys_model}.
Due to the high atmospheric absorption associated with mmW signals, the radar antenna beams are narrow and we assume that the antenna gain, $G$, is uniform within the main lobe and inversely proportional to beamwidth. Thus $G = G_0\Delta\psi^{-1}$ where $G_0$ is a constant of proportionality. For mathematical simplicity, we consider a scenario with a single target located at a distance $r_t$ from the radar, which is within its main lobe and whose radar cross-section, $\sigma_t$, follows the exponential/Swerling-1 model with mean RCS $\sigma_{t_{avg}}$. Based on the radar range equation, the received radar signal is
\par\noindent \small
\begin{align}
\label{equ:rx_sig}
  \mathbf{S}(r_t) =P_{tx}G^2\mathbf{\sigma_t} r_t^{-2\alpha} = P_0 \mathbf{\sigma_t}. r_t^{-2\alpha}.
\end{align} \normalsize
Here, the two-way mmW propagation is modeled by the inverse $\alpha$-law as given in \cite{al2014propagation}. All the constants in the above expression are combined into a single constant $P_0$.

Next, we consider the returns from clutter scatterers within the same resolution cell ($A_r$) as the target. The discrete clutter scatterers are modeled as a Poisson point process (PPP) distribution, $\Phi_c$ within the radar field-of-view, wherein the number of clutter scatterers in each realization of the spatial stochastic process follows a Poisson distribution while the positions follow a uniform distribution. 
In a PPP, each point is stochastically independent of all the other points in the process. We model the Rayleigh fading process in the channel as given in \cite{park2018analysis} using $\mathbf{g_c}$, a random variable with an exponential distribution with mean $g_{c_{avg}}$. The RCS of clutter scatterers is represented as ${\sigma_c}$. Variation in the strength of clutter returns is modeled by the clutter density $\rho_c$ in $\Phi_c$. Now, the clutter returns at the radar receiver is modeled as - 
\par\noindent\small
\begin{align}
\label{equ:rx_clutter}
  \mathbf{C} = \sum_{c\in \Phi_c,A_r} P_{tx}G^2\mathbf{g_c}\sigma_c r_c^{-2\alpha} = \sum_{c\in \Phi_c, A_r} P_0\mathbf{g_c}\sigma_c r_c^{-2\alpha}.
\end{align} \normalsize
The radar receiver is also characterized by thermal noise with mean noise power as $N_p = K T_s BW$, where $K$ is the Boltzmann constant, $T_s$ is the system temperature, and $BW$ is the bandwidth of the radar signal.
 \section{Estimation of Radar Operating Metrics}
\label{sec:ROM}
In this section, we discuss the radar operating metrics which include the probability of detection ($\mathcal{P}_{d}$) and false alarms ($\mathcal{P}_{fa}$) for an urban mmW environment with target and clutter. 
 \subsection{Probability of False Alarm}
 $\mathcal{P}_{fa}$ is defined as the probability that the received signal power exceeds the pre-defined threshold $\eta$ in the absence of the target, as shown here
 \par\noindent\small
\begin{align}
\label{eq:p_fa1}
  \mathcal{P}_{fa} = \mathcal{P}( \mathbf{C}+N_p \geq \eta) = 1-\mathcal{P}( \mathbf{C} \leq \eta - N_p).
\end{align} \normalsize
Thus, we can estimate $\mathcal{P}_{fa}$ by calculating cumulative distribution function (CDF) of clutter $\mathcal{F}(\mathbf{C})$ as given
 \par\noindent\small
\begin{align}
\label{eq:p_fa}
  \mathcal{P}(\mathbf{C} \leq \eta - N_p) = \mathcal{F}(\mathbf{C}) = \mathcal{F}(\eta - N_p).
\end{align} \normalsize
The CDF of a function $f(x)$ with random variable $\mathbf{x}$ is obtained using the  Gil-Pelaez’s inversion theorem \cite{gil1951note} as - 
\par\noindent\small
\begin{align}
  \begin{aligned}
\mathcal{F}(\mathbf{x})=\frac{1}{2}- \frac{1}{\pi}\int_{0}^{\infty}\frac{1}{\omega} Im\left[\varphi_\mathbf{x}(\omega)exp(-j\omega \mathbf{x})\right]d\omega, 
\end{aligned}
\end{align}\normalsize
where $\varphi_\mathbf{x}(\omega)$ is the characteristic function (CF) that is defined as $\varphi_\mathbf{x}(\omega) = \mathbb{E}\left[e^{j\omega \mathbf{x}}\right]$. Hence, we first need to obtain the CF of the clutter model. Note that the clutter returns are a function of random variables which include the position of the clutter scatterers and $\mathbf{g}_c$. Now, CF of the clutter is expressed as -
\par\noindent\small
\begin{align}
\varphi_\mathbf{c}(\omega) = \mathbb{E}_{g_c} \mathbb{E}_{\Phi_c}\left[\exp \left(j \omega P_0\sigma_c\sum_{c\in \Phi_c,A_r} \mathbf{g}_cr_c^{-2\alpha}\right)\right]. 
\end{align}\normalsize
Here,  $\mathbb{E}_{\Phi_c}$ signifies the expectation operator for the positions clutter scatterers, and $\mathbb{E}_{g_c}$ signifies the expectation operator for the random variable $\mathbf{g}_c$. Since the exponent of the sum of terms can be written
as a product of exponents, we write the above expression as - 
\par\noindent\small
\begin{align}
\label{equ:exp_eq}
\varphi_\mathbf{c}(\omega) =\mathbb{E}_{\Phi_c,g_c}\left[\prod_{\Phi_c,A_r} \exp \left(j \omega P_0 \sigma_c\mathbf{g}_cr_c^{-2\alpha}\right)\right].
\end{align}\normalsize
Now, we apply the probability-generating
functional (PGFL) of the homogeneous PPP \cite{haenggi2012stochastic}, which is expressed as 
\par\noindent\small
\begin{align}
\begin{aligned}
 E\left[\prod_{\Phi} f(x)\right] =\exp \left[-\iint_{A_r}(1-f(x) \rho(d \mathbf{x})\right].
\end{aligned}
\end{align}\normalsize
Further, the two-dimensional region of interest in our model is the area of the range-azimuth cell $A_r$ corresponding to $r_c$. $A_r$ is a function of $\Delta\psi$ and range resolution $\Delta r$ as shown in
\par\noindent\small
\begin{align}
\begin{aligned}
 A_r &= r_c \Delta\psi \Delta r 
 = \frac{r_c \Delta\psi c\tau}{2},
\end{aligned}
\end{align}\normalsize
where, $c$ and $\tau$ denote the speed of light and pulse width of the signal, respectively. The clutter is assumed to be uniform within $A_r$.
Thus, Eq.~\ref{equ:exp_eq} is written as
\par\noindent\small
\begin{align}
\varphi_\mathbf{c}(\omega)  =\exp \left[-\rho_c A_r \left(\mathbb{E}{_{g_c}(1-\exp (j \omega P_0  \sigma_c\mathbf{g_c} r_c}^{-2 \alpha}))\right)\right] \\
=\exp \left(-\rho_c A_r \int_0^{\infty}\left(1-\exp \left(j \omega P_0 \sigma_c\mathbf{g}_c r_c^{-2 \alpha}\right)\right) \mathcal{P}\left(g_c\right) d g_c\right)  
\end{align}\normalsize

On solving the above integration, we obtain the characteristic function of clutter as 
\par\noindent\small
\begin{align}
\begin{aligned}
\varphi_\mathbf{c}(\omega)=\exp \left(-\rho_c A_r \left(\frac{j \omega P_0r_c^{-2 \alpha}\sigma_c g_{c_{avg}}}{j \omega P_0 r_c^{-2 \alpha} \sigma_cg_{c_{avg}}-1}\right)\right)
.
\end{aligned}
\end{align}\normalsize
Now, the CDF of clutter is obtained using Gil-Pelaez’s inversion theorem as shown  
 \par\noindent\small
\begin{align}
\label{eq:cdf_clutter1}
\mathcal{F}(\mathbf{C})=\frac{1}{2}-\frac{1}{\pi} \int_0^{\infty}\frac { 1 } { \omega } I_{ m }  \Biggl[ \operatorname { e x p } \Biggl(-\rho_c A_r \nonumber \\ \left(\frac{j \omega P_0 r_c^{-2 \alpha}\sigma_c g_{c_{avg}}}{j \omega P_0 r_c^{-2 \alpha} \sigma_cg_{c_{avg}}-1}\right)\Biggr) \exp (-j \omega\mathbf{C})\Biggr] d \omega.
\end{align}\normalsize
We combine $\sigma_c$ and $\rho_c$ to model the surface clutter coefficient $\sigma_o$ such that
$\sigma_o = \sigma_c\rho_c$ as given in \cite{skolnik1980introduction}. Thus, the above equation is written as 
\par\noindent\small
\begin{align}
\label{eq:cdf_clutter}
\mathcal{F}(\mathbf{C})=\frac{1}{2}-\frac{1}{\pi} \int_0^{\infty}\frac { 1 } { \omega } I_{ m }  \Biggl[ \operatorname { e x p } \Biggl(-\sigma_o A_r \nonumber \\ \left(\frac{j \omega P_0 r_c^{-2 \alpha} g_{c_{avg}}}{j \omega P_0 r_c^{-2 \alpha} \sigma_cg_{c_{avg}}-1}\right)\Biggr) \exp (-j \omega\mathbf{C})\Biggr] d \omega.
\end{align}\normalsize
We now substitute the value of $\mathbf{C}$ from Eq.~\ref{eq:p_fa} in the above equation to obtain $\mathcal{P}_{fa}$ as shown in
\par\noindent\small
\begin{multline}
\label{eq:cdf_clutter_final}
\mathcal{P}_{fa} = 1-\mathcal{F}(\mathbf{C}) =\frac{1}{2}+\frac{1}{\pi} \int_0^{\infty}\frac { 1 } { \omega } I_{ m} (\operatorname { e x p } \Biggl(-\sigma_o A_r \\
\Biggl(\frac{j \omega P_0 r_c^{-2 \alpha} g_{c_{avg}}}{j \omega P_0 r_c^{-2 \alpha} \sigma_cg_{c_{avg}}-1}\Biggr)\Biggr) \exp (-j \omega(\eta-N_p))) d \omega.
\end{multline}\normalsize
Since the above equation cannot be solved analytically, we use numerical integration to obtain $\mathcal{P}_{fa}$. 

\subsection{Probability of Detection}
The probability of detection ($\mathcal{P}_{d}$) is defined as the probability that the
envelope of the signal received at the radar receiver (which includes target returns along with clutter and noise) exceeds a pre-defined threshold value $\eta$ as shown here
 \par\noindent\small
\begin{align}
\label{eq:p_d}
  \mathcal{P}_{d} = \mathcal{P}(  \mathcal{S}+\mathbf{C}+N_p \geq \eta) = 1-\mathcal{P}( \mathbf{C} \leq \eta - N_p-S_0),
\end{align} \normalsize
where $S_0$ is the mean received signal power. The value of $S_0$ is obtained by taking the expectation of Eq.~\eqref{equ:rx_sig} for the Swerling distribution of $\sigma_t$.
Now, we determine $\mathcal{P}_{d}$ by calculating the CDF of clutter $\mathcal{F}(\mathbf{C})$ by 
 \par\noindent\small
\begin{align}
\label{eq:p_d2}
\mathcal{F}(\mathbf{C}) = \mathcal{F}( \eta - N_p -S_0).
\end{align} \normalsize
Here, $\mathcal{F}(\mathbf{C})$ is directly obtained form Eq.~\ref{eq:cdf_clutter_final}. Thus, the final expression for $\mathcal{P}_{d}$ is 
\par\noindent\small
\begin{multline}
\label{eq:pd_clutter_final}
\mathcal{P}_{d} = \frac{1}{2}+\frac{1}{\pi} \int_0^{\infty}\frac { 1 } { \omega } I _ { m } (\operatorname { e x p } \Biggl(-\sigma_o A_r \\
\Biggl(\frac{j \omega P_0 r_c^{-2 \alpha} g_{c_{avg}}}{j \omega P_0 r_c^{-2 \alpha} \sigma_cg_{c_{avg}}-1}\Biggr)\Biggr) \exp (-j \omega(\eta-N_p-S_0))) d \omega,
\end{multline}\normalsize
which is obtained using numerical integration.
\section{ISAC System Model for Throughput Estimation}
\label{sec:System Model}
\begin{figure}[htbp]
    \centering   
    \includegraphics[scale=0.5]{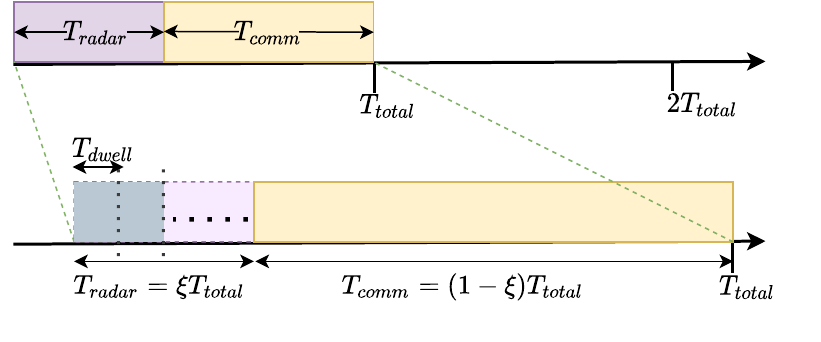}
    \caption{Time-multiplexing of radar and communication functionalities within an ISAC framework.}
    \label{fig:beam_scan}
\end{figure}

In 802.11ad-based ISAC systems, radar and communication functionalities share the spectrum and hardware resources in a time-multiplexed manner as shown in Fig.\ref{fig:beam_scan}. Here, the total time interval $T_{total}$ per cycle of ISAC operation is split between  $T_{radar} = \xi T_{total}$ for radar and $T_{comm} = (1-\xi)T_{total}$ where $\xi$ is the radar duty cycle.
We assume that $T_{radar}$ time interval comprises multiple $T_{dwell}$ which corresponds to the dwell/scan time of the radar antenna beam of $\Delta\psi$ beamwidth. 
Within the $T_{dwell}$ interval, the radar is assumed to transmit multiple pulses for range and Doppler estimation. For a fixed $T_{dwell}$, the radar beamwidth will have to be adjusted based on the available time for radar ($\xi T_{total}$). In other words, if $T_{radar}$ is long, then very narrow beams of high gain are used to scan the angular search space $\Omega$. On the other hand, if $T_{radar}$ is short, then broader beams of lower gain have to be used to scan the same $\Omega$. Thus, $\Delta\psi$ is inversely proportional to $\xi$ as shown in
\par\noindent\small
\begin{align}
\label{equ:beam_scan}
 {\Delta\psi} = \frac{T_{dwell} \Omega}{\xi T_{total} }.
\end{align}
 \normalsize
Hence, the term $P_0$ in the target scattered signal modeled in Eq.~\ref{equ:rx_sig} and in the clutter returns model in Eq.~\ref{equ:rx_clutter} are now functions of $\xi$. As a result, for a smaller $\xi$, we get lower $\mathcal{P}_d$ and fewer targets are likely to get detected. 

The average number of detected targets for a specific $\xi$ is given by
$\beta(\xi)$. If we assume a uniform distribution of the mobile targets of spatial density $\rho_t$, then 
\par\noindent\small
\begin{align}
\beta(\xi) = \mathcal{P}_d \rho_t\Omega r_t c\tau/2,
\end{align}
\normalsize
where $c\tau/2$ is the range resolution.
Note that in the above discussion, clutter scatterers are also likely to be detected by the radar but will ultimately be ignored by the communication modules due to the lack of acknowledgment of downlink pilots. Then the overall network throughput ($\gamma$) for a communication service time of $(1-\xi)T_{total}$ is 
\par\noindent\small
\begin{align}
  \gamma & = \beta(\xi)(1-\xi)D.
\end{align} \normalsize
Here, we assume that the communication resources are available to all the detected targets and that they are all serviced by identical communication data rates $D$. Thus we see that $\gamma$ is a function of $\beta(\xi)$ which increases with $\xi$ as well as $T_{comm}$ which reduces with $\xi$.
\section{Numerical Results}
\label{sec: Numerical Results}
In this section, we perform simulations to analyze the effect of radar, target, and clutter parameters on  $\mathcal{P}_{fa}$ and communication link metrics in terms of $\gamma$. The parameters chosen to carry out the simulation are given in Table~\ref{tab:sim_parameters}.
\begin{table}[htbp]
 \caption{ Simulation Parameters}
    \centering
    \begin{tabular}{c|c|c}
   \hline\hline
    \textbf{Parameters}& \textbf{Symbols}& \textbf{Values} \\
    \hline\hline
    Carrier frequency &  $f_c$  & 60 GHz\\
    Bandwidth & $BW$  & 20 MHz \\
     Transmitted Power &  $P_{tx}$  & 1 W\\  
       Gain constant& $G_o$ & 1\\
      RCS of clutter& $\sigma_{c}$ & 0.1 $m^2$\\
       Mean RCS of target& $\sigma_{t}$ & 10 $m^2$ \\
        Clutter density& $\rho_c$  & 1 $m^{-2}$ \\
        Fading coefficient & $g_{c_{avg}}$  & 1 \\
        Threshold & $\eta$  & 1e-13 W \\
        Range of clutter & $r_c$  & 10 m \\
        Range of target & $r_t$  & 10 m \\
        Standard temperature & $T_s$  & 300 K \\
        Duty Cycle & $\xi$  & 0.9 \\
       \hline
    \end{tabular}
    \label{tab:sim_parameters}
\end{table}
\subsection{Effect of transmitted power on $\mathcal{P}_{fa}$ and $\gamma$ }
\begin{figure}[htbp]
    \centering   
    \includegraphics[scale = 0.24]{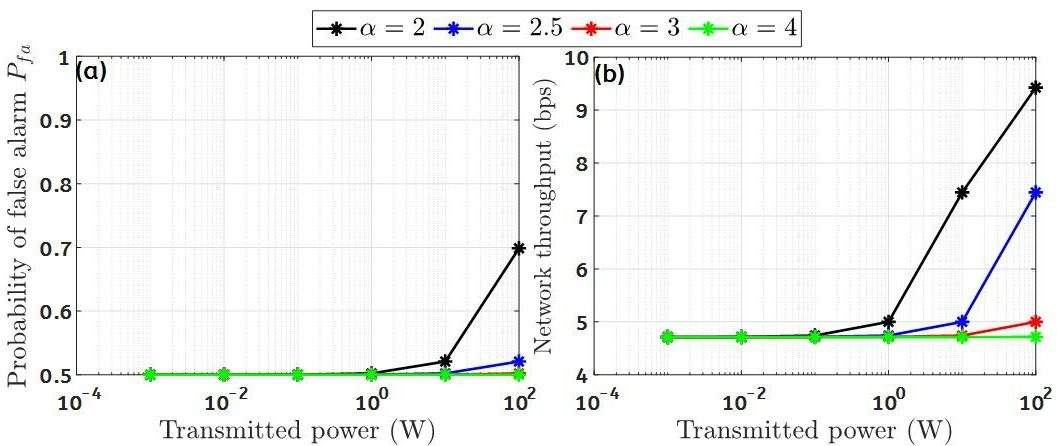}
    \caption{ Probability of false alarm and network throughput for different transmitted powers.}
    \label{fig:pd_pfa_vs_ptx}
\end{figure}

.
In Fig.~\ref{fig:pd_pfa_vs_ptx}(a), we compare $\mathcal{P}_{fa}$ for different values of transmitted power $P_{tx}$ and for different channel conditions by varying $\alpha$ from 2 to 4. Note that a higher value of $\alpha$ denotes a deteriorated channel with high attenuation. The rest of the parameters are considered the same as reported in Table~\ref{tab:sim_parameters}. We observe that $\mathcal{P}_{fa}$ increases with the increased transmitted power since the clutter returns increase, resulting in higher false alarms. Also, we observe that there is a significant degradation in $\mathcal{P}_{fa}$ with the increase in $\alpha$.
Next, we analyze $\gamma$ for different values of transmitted power in Fig.~\ref{fig:pd_pfa_vs_ptx}(b), keeping the rest of the parameters the same as reported in Table~\ref{tab:sim_parameters}. Here, we observe that $\gamma$ increases with the increase in transmitted power due to improvement in the probability of detection of multiple targets.
\subsection{Effect of bandwidth on $\mathcal{P}_{fa}$ and $\gamma$ }
\begin{figure}[htbp]
    \centering   
    \includegraphics[scale = 0.23]{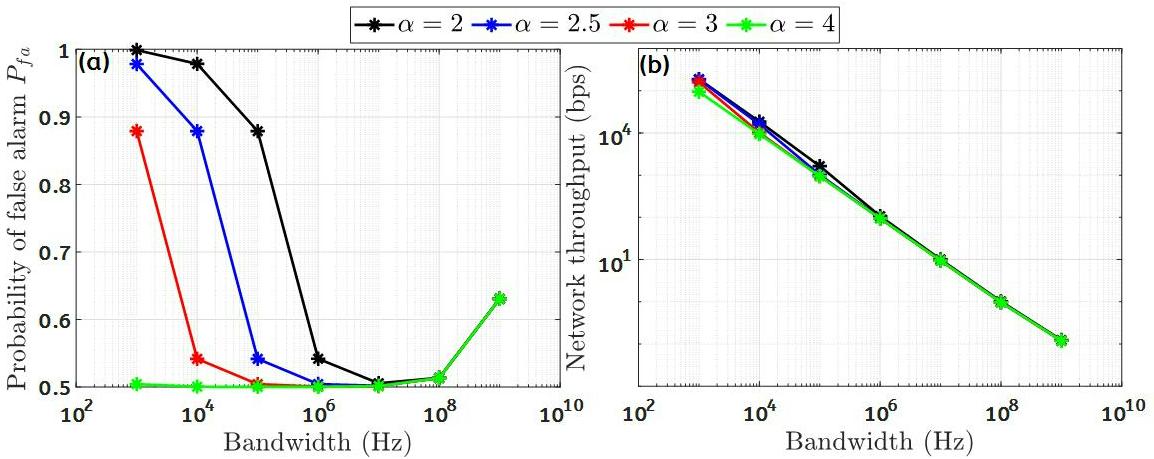}
    \caption{ Probability of false alarm and network throughput for different bandwidths.}
    \label{fig:pd_pfa_vs_bw}
\end{figure}

We analyze the $\mathcal{P}_{fa}$ and $\gamma$ for different bandwidths in Fig.~\ref{fig:pd_pfa_vs_bw}. $\mathcal{P}_{fa}$ decreases initially with increased bandwidth. This is because the clutter area $A_r$ is inversely proportional to the bandwidth of the radar signal. However, $\mathcal{P}_{fa}$ increases after a bandwidth of 100 MHz because the noise power after this bandwidth is very high. Again, a fall in $\mathcal{P}_{fa}$ is observed with an increase in $\alpha$.
Also, $\gamma$ is significantly high (up to 0.1 Mbps) for a lower bandwidth as is observed in Fig.~\ref{fig:pd_pfa_vs_bw}(b). However, $\gamma$ decreases linearly with the increase in bandwidth. This surprising result is because the data rates are kept fixed (and do not scale with bandwidth in our model). However, the number of detected targets is reduced due to high noise power resulting in overall degradation of throughput. 

\subsection{Effect of duty cycle on $\mathcal{P}_{fa}$ and $\gamma$}
\begin{figure}[htbp]
    \centering   
    \includegraphics[scale = 0.23]{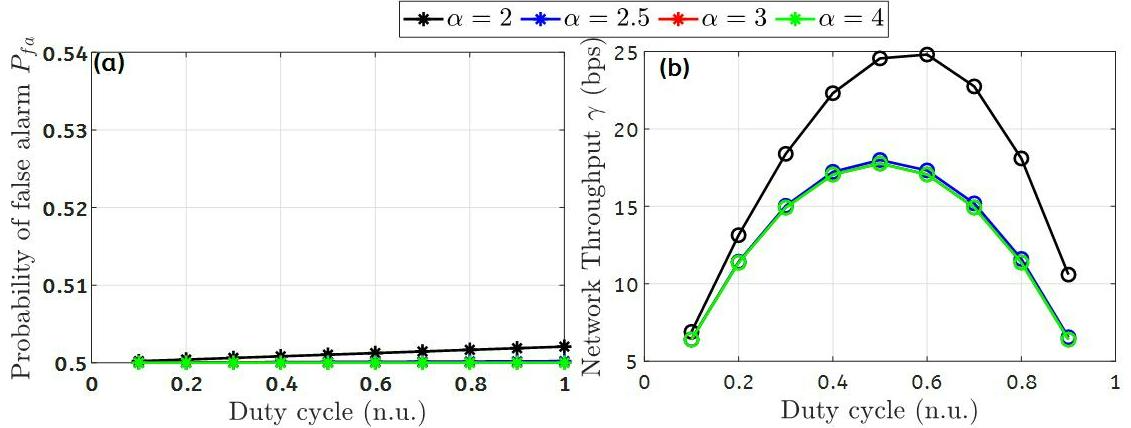}
    \caption{Probability of false alarm and network throughput for different duty cycles.}
    \label{fig:pd_pfa_vs_epsilon}
\end{figure}

Here, we analyze $\mathcal{P}_{fa}$ and $\gamma$ for duty cycle varying from 0 to 1 in Fig.~\ref{fig:pd_pfa_vs_epsilon}. As $T_{radar}$ time is increased, for higher $\xi$, more duration of $T_{total}$ is allotted for the radar search. This results in more false alarms due to higher antenna gain. Hence, we observe that $\mathcal{P}_{fa}$ increases with the duty cycle. However, there is slight or no impact of the duty cycle for higher $\alpha$ on $\mathcal{P}_{fa}$ as clutter power has reduced significantly due to the attenuation of the signals. 
In Fig.~\ref{fig:pd_pfa_vs_epsilon}(b), we discuss the performance of $\gamma$ with respect to $\xi$. We observe that $\gamma$ initially increases with an increase in $\xi$ and a corresponding increase in $\mathcal{P}_{d}$. But subsequently, $\gamma$ deteriorates due to the reduction in communication service time.  Also, for different values of $\alpha$, the variation of $\gamma$ with respect to $\xi$ is similar.

\subsection{Effect of surface clutter coefficient on $\mathcal{P}_{fa}$ and $\gamma$}
\begin{figure}[htbp]
    \centering   
    \includegraphics[scale = 0.23]{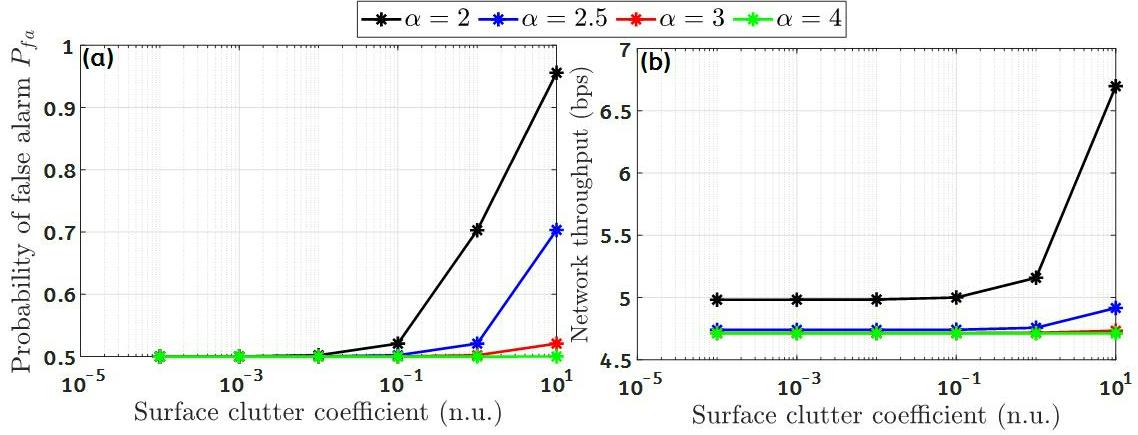}
    \caption{Probability of false alarm and throughput for different surface clutter coefficients.}
    \label{fig:pd_pfa_vs_rho}
\end{figure}

In Fig.~\ref{fig:pd_pfa_vs_rho}, we vary the surface clutter coefficient $\sigma_o$  and analyze $\mathcal{P}_{fa}$ and $\gamma$. With increased $\sigma_o$, the clutter power increases, resulting in an increase in $\mathcal{P}_{fa}$. Thus, we observe in Fig.~\ref{fig:pd_pfa_vs_rho}(a), $\mathcal{P}_{fa}$ reach up to 100\%  for a very high $\sigma_o$ at lower $\alpha$. However, for large values of $\alpha$, there is no change with respect to $\sigma_o$.
We analyze the performance of $\gamma$ with respect to $\sigma_o$ in Fig.~\ref{fig:pd_pfa_vs_rho}(b) and observe that $\gamma$ increases with increase in $\sigma_o$. Further, $\gamma$ degrades for higher $\alpha$.  

\subsection{Effect of RCS of target on $\mathcal{P}_{fa}$ and $\gamma$}
\begin{figure}[htbp]
    \centering   
    \includegraphics[scale = 0.23]{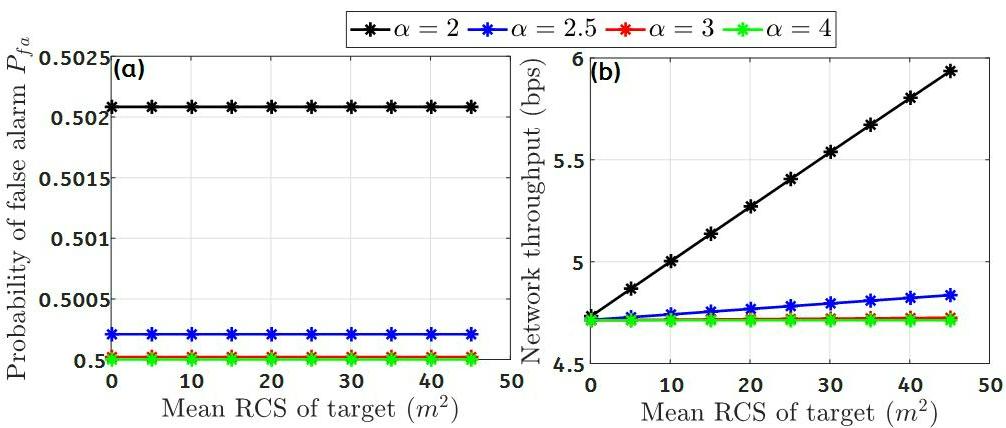}
    \caption{ Probability of false alarm and network throughput for different RCS of the target.}
    \label{fig:pd_pfa_vs_rcs_mu}
\end{figure}

In Fig.~\ref{fig:pd_pfa_vs_rcs_mu}(a), we compare $\mathcal{P}_{fa}$ with respect to different values of RCS of the target $\sigma_t$. We observe that there is no change in $\mathcal{P}_{fa}$ for different values of  $\sigma_t$ as  $\mathcal{P}_{fa}$ does not depend on the received signal from the target. Further, we observe in  Fig.~\ref{fig:pd_pfa_vs_rcs_mu}(b), that there is an increment in the $\gamma$ as the value of $\sigma_t$ increases. This is because a higher value of $\sigma_t$ increases the received signal power, leading to a higher probability of detection.
\section{Conclusion}
\label{sec:Conclusion}
We estimate the radar operating metrics and network throughput of the ISAC framework under noise and clutter-limited conditions in complex propagation environments using stochastic geometry. Further, we analyze the effect of different parameters on the probability of false alarms and network throughput. 
Note that in this work, we consider a fixed threshold to estimate the probability of detection. In our revised work, we will extend this framework to determine the optimum threshold for maximizing the probability of detection for a constant probability of false alarms.

\bibliographystyle{ieeetran}
\bibliography{reference}
\end{document}